\documentclass[aps,10pt,twocolumn,showpacs,prb,superscriptaddress]{revtex4-1}
\usepackage{graphicx}
\usepackage{dcolumn}
\usepackage{amsmath}
\usepackage{amssymb}
\usepackage{accents}
\usepackage{soul}

\renewcommand{\Re}{\mathop\mathrm{Re}\nolimits}

 \DeclareMathOperator{\Tr}{Tr}

\makeatletter
\def\btt#1{\texttt{\@backslashchar#1}}
\DeclareRobustCommand\bblash{\btt{\@backslashchar}} \makeatother

\begin{document}

\title{Consequences of bulk odd-frequency superconducting states\\ for the classification of Cooper pairs}

\author{Yasuhiro Asano}
\affiliation{Department of Applied Physics and Center for Topological Science \& Technology,
Hokkaido University, Sapporo 060-8628, Japan}
\affiliation{Moscow Institute of Physics and Technology, 141700 Dolgoprudny, Russia}

\author{Yakov V.\ Fominov}
\affiliation{L.~D.\ Landau Institute for Theoretical Physics, RAS, 142432, Chernogolovka, Russia}
\affiliation{Moscow Institute of Physics and Technology, 141700 Dolgoprudny, Russia}

\author{Yukio Tanaka}
\affiliation{Department of Applied Physics, Nagoya University, Nagoya 464-8603, Japan}
\affiliation{Moscow Institute of Physics and Technology, 141700 Dolgoprudny, Russia}

\date{2 November 2014}

\begin{abstract}
We analyze symmetries and magnetic properties of Cooper pairs appearing
as subdominant pairing correlations in inhomogeneous superconductors,
on the basis of the quasiclassical Green-function theory.
The frequency symmetry, parity, and the type of magnetic response of such subdominant
correlations are opposite to those of the dominant pairing correlations in the bulk state.
Our conclusion is valid even when we generalize the theory of superconductivity
to recently proposed diamagnetic odd-frequency superconductors.
As a consequence, Cooper pairs are classified into eight classes
in terms of their symmetries and magnetic properties.
Anomalous magnetic properties of subdominant components can be probed by studying the Meissner effect.
\end{abstract}


\maketitle

\section{Introduction}

Superconductors demonstrate the diamagnetic response to magnetic field
below the transition temperature as a result of
spontaneous breaking of the continuous gauge symmetry.
At the same time, the superconductor
chooses three discrete symmetry options for Cooper pairing: frequency symmetry, spin configuration,
and parity. In each case there are two options: pairing can be either
symmetric or antisymmetric with respect to interchange of the corresponding
arguments: times, spins, or coordinates of two electrons forming a Cooper pair.
Since electrons obey the Fermi-Dirac statistics, there is the constraint that the pairing functions must be
antisymmetric under permutation of the two electrons (i.e., simultaneous permutation of all the arguments).
As a consequence, Cooper pairs has been classified into four symmetry classes.
The pair potentials in all superconductors discovered so far belong to
the even-frequency symmetry class.
A number of theories have suggested the appearance of Cooper pairs (superconducting correlations)
belonging to the odd-frequency symmetry class in superconducting proximity structures.
Spatial inhomogeneities such as surfaces and interfaces break
the translational symmetry, thus leading to the coexistence of even-parity and
odd-parity Cooper pairs.\cite{tanaka07,eschrig2} Spin-dependent potentials\cite{bergeret,ya07sfs,braude,eschrig}
enable mixing of spin-singlet and spin-triplet pairs.
In such inhomogeneous superconducting structures, the odd-frequency Cooper pairs appear
as a subdominant component of the pairing correlations.
As a consequence, anomalous low energy transport due to the odd-frequency pairs have been reported in
superconducting junctions.\cite{yt04,ya06,ya07,fominov}

In addition to the pairing symmetry, Cooper pairs are also characterized by their magnetic properties.
The even-frequency pairs in the bulk are usually diamagnetic,
which corresponds to conventional positive pair density. Recent studies have suggested that the subdominant odd-frequency Cooper pairs formally have \emph{locally negative pair density} which physically signifies unconventional paramagnetic response to the magnetic field.\cite{yt05r,ya11,higashitani1,mironov,suzuki}
The odd-frequency pairs, however, are not necessarily paramagnetic.
Originally, the odd-frequency pairs were discussed in the framework of odd-frequency
superfluidity\cite{berezinskii} and
superconductivity.\cite{balatsky92,kirkpatrik,vojta,coleman,solenov,kusunose}
Diamagnetic odd-frequency pairs could form a homogeneous superconducting ground state.\cite{kirkpatrik,solenov,kusunose}
Although several theoretical papers have suggested possibilities of odd-frequency superconductivity
in strongly correlated electron systems~\cite{fuseya,kusunose2,hoshino},
 no clear experimental evidence of odd-frequency superconductivity has been presented so far.
It is difficult to resolve directly the frequency symmetry of pair potentials.
This is part of the reason for lacking experimental confirmation.
The theoretical prediction of the characteristic phenomena in such exotic phase could assist
the detection of the odd-frequency superconductivity.
In this paper, for this purpose, we generalize the theory of superconductivity to odd-frequency superconductors.
If we assume the existence of an odd-frequency superconductor, then an inhomogeneity can generate subdominant Cooper pairs.
By applying the generalized theory, we study symmetries and magnetic properties of such subdominant Cooper pairs.

Technically, we derive the quasiclassical Eilenberger equation\cite{eilenberger,larkin} in the form that
can be applied, in particular, to the odd-frequency superconductors.
The inhomogeneity in superconducting pair potential generates
the subdominant Cooper pairs which have the opposite parity,
the opposite frequency symmetry, and the opposite type of magnetic response, compared to those of the
dominant Cooper pairs in the bulk (at the same time, the spin symmetry is preserved since we do not consider spin-dependent potentials).
Our results imply the ubiquitous presence of the paramagnetic Cooper pairs.
The subdominant Cooper pairs behave as if recovering
the broken discrete symmetry options for Cooper pairing and the global gauge symmetry.
We conclude that the assumption of possibility of bulk odd-frequency superconducting states leads to a generalized classification of Cooper pairs into eight classes in terms of
their pairing symmetries and magnetic properties.
In the end of this paper, we discuss in more detail the special case of
subdominant Cooper pairs which belong to the conventional
even-frequency spin-singlet $s$-wave symmetry class but at the same time are paramagnetic.

\section{Mean-field theory}
Throughout this paper, we repeatedly use three sign factors representing
the symmetries of the pair potential,
\begin{align}
s_f &=\left\{\begin{array}{cc}
   1 &  \text{even-frequency}\\
  -1 &  \text{odd-frequency}\end{array}\right.,\\
s_p &=\left\{\begin{array}{cc}
   1 &  \text{even-parity}\\
  -1 &  \text{odd-parity}\end{array}\right.,\\
s_s &=\left\{\begin{array}{cc}
   1 &  \text{spin-triplet}\\
  -1 &  \text{spin-singlet}\end{array}\right..
\end{align}
The Fermi-Dirac statistics of electrons imply $s_f s_p s_s =-1$.
The Pauli matrices are denoted by $\hat{\sigma}_j$ for $j=1,2,3$.
The unity matrix in the spin space is denoted by $\hat{\sigma}_0$.
We use the units of $\hbar=k_\mathrm{B}=c=1$, where $k_\mathrm{B}$ is the Boltzmann constant and $c$
is the speed of light.

\subsection{Gor'kov equation}

In the mean-field theory of superconductivity, two kinds of the pair potentials
appear in the Gor'kov equation,
\begin{align}
\int & d\mathrm{x}_1 \left\{
 -\delta(\mathrm{x}-\mathrm{x}_1) \partial_\tau \check{T}_3
 -\left[ \begin{array}{cc}
 \hat{h}(\mathrm{x},\mathrm{x}_1)
 &  \hat{\Delta}(\mathrm{x},\mathrm{x}_1)\\
-\underline{\hat{\Delta}}(\mathrm{x},\mathrm{x}_1) &
 \hat{h}^\ast(\mathrm{x},\mathrm{x}_1)
 \end{array}
 \right]
\right\} \notag \\
&\times
\left[\begin{array}{cc}
\hat{\mathfrak{G}}(\mathrm{x}_1,\mathrm{x}') & \hat{\mathfrak{F}}(\mathrm{x}_1,\mathrm{x}') \\
 & \\
-\underline{\hat{\mathfrak{F}}}(\mathrm{x}_1,\mathrm{x}') & -\underline{\hat{\mathfrak{G}}}(\mathrm{x}_1,\mathrm{x}')
\end{array}\right]
= \check{1}
\delta(\mathrm{x}-\mathrm{x}'), \label{gorkov_l}\\
&\hat{h}(\mathrm{x},\mathrm{x}_1)= \delta(\mathrm{x}-\mathrm{x}_1)
\left[ - \frac{\left( \nabla_1 -{ie}\boldsymbol{A}(\boldsymbol{r}_1)\right)^2}{2m}
  -\mu_F \right]\hat{\sigma}_0, \notag \\
&\quad \check{T}_3 =\left[\begin{array}{cc}
\hat{\sigma}_0 & \hat{0} \\
\hat{0} & -\hat{\sigma}_0
\end{array}\right],
\end{align}
with $\mathrm{x}=(\boldsymbol{r},\tau)$ in the imaginary-time representation and $\check{1}=\text{diag}[\hat{\sigma}_0,\hat{\sigma}_0]$.
The pair potentials enter the self-consistency equations
\begin{align}
\Delta_{\alpha,\beta}(\mathrm{x},\mathrm{x}') &= \sum_{\gamma,\delta}
V_{\alpha\beta; \gamma\delta}(\mathrm{x},\mathrm{x}')
{\mathfrak{F}}_{\gamma,\delta}(\mathrm{x},\mathrm{x}'),\label{delta_def1}
\\
\underline{\Delta}_{\alpha,\beta}(\mathrm{x},\mathrm{x}') &=
\sum_{\gamma,\delta} V^\ast_{\alpha\beta; \gamma\delta}(\mathrm{x},\mathrm{x}')
\underline{{\mathfrak{F}}}_{\gamma,\delta}(\mathrm{x},\mathrm{x}'),\label{deltatilde_def1}
\end{align}
 where $V_{\alpha\beta; \gamma\delta}(\mathrm{x},\mathrm{x}')$
represents the pairing interaction (spin-dependent in the general case, hence the subscript).
The four electron Green functions are defined by
\begin{align}
\mathfrak{G}_{\alpha,\beta}(\mathrm{x},\mathrm{x}')&=-\bigl< T_\tau
\psi_\alpha(\mathrm{x}) \psi_\beta^\dagger(\mathrm{x}')\bigr>,\\
\underline{\mathfrak{G}}_{\alpha,\beta}(\mathrm{x},\mathrm{x}')&=-\bigl< T_\tau
\psi_\alpha^\dagger(\mathrm{x}) \psi_\beta(\mathrm{x}')\bigr>,\\
\mathfrak{F}_{\alpha,\beta}(\mathrm{x},\mathrm{x}')&=-\bigl< T_\tau
\psi_\alpha(\mathrm{x}) \psi_\beta(\mathrm{x}')\bigr>,\\
\underline{\mathfrak{F}}_{\alpha,\beta}(\mathrm{x},\mathrm{x}')&=-\bigl< T_\tau
\psi_\alpha^\dagger(\mathrm{x}) \psi_\beta^\dagger(\mathrm{x}')\bigr>. \label{def_gorkov_m}
\end{align}
The functional-integral theories\cite{solenov,kusunose} have suggested that a saddle point solution
\begin{equation}
\underline{\Delta}_{\alpha,\beta}(\mathrm{x},\mathrm{x}')=
-\Delta_{\alpha,\beta}^\ast(\mathrm{x},\mathrm{x}'),\label{choice}
\end{equation}
minimizes the free energy and describes
the uniform diamagnetic superconducting ground states for any frequency symmetry.\cite{kusunose}
Indeed, the usual equal-time pair potentials satisfy Eq.\ (\ref{choice}).
The pair potential is expressed in the Fourier
representation and decomposed into spin components as
\begin{align}
\hat{\Delta}(\mathrm{x}_1,\mathrm{x}_2) &= \hat{\Delta}(\boldsymbol{r}_{12},\boldsymbol{\rho}_{12},\tau_{12}),\\
=\int\frac{d\boldsymbol{k}}{(2\pi)^d} &T\sum_{\omega_n}
  \hat{\Delta}(\boldsymbol{r}_{12},\hat{\boldsymbol{k}},i\omega_n)
e^{i \boldsymbol{k}\cdot \boldsymbol{\rho}_{12}}
e^{-i \omega_n \tau_{12}},\\
\hat{\Delta}(\boldsymbol{r},\hat{\boldsymbol{k}},i\omega_n)&= \sum_{\nu=0}^{3}
\Delta_\nu(\boldsymbol{r},\hat{\boldsymbol{k}},i\omega_n) i \hat{\sigma}_\nu
 \hat{\sigma}_2  \;e^{i\varphi}, \label{delta0}
\end{align}
where $d$ is the dimensionality of the superconductor, $\varphi$ is the superconducting phase,
and $\omega_n=(2n+1)\pi T$ is the
fermionic Matsubara frequency with $n$ being an integer number and $T$ being the temperature.
The center-of-mass coordinate is denoted by $\boldsymbol{r}_{12}= (\boldsymbol{r}_1+\boldsymbol{r}_2)/2$.
The relative coordinates are represented by $\boldsymbol{\rho}_{12}=\boldsymbol{r}_1-\boldsymbol{r}_2$,
and $\tau_{12}=\tau_1-\tau_2$.
In the weak-coupling limit, the orbital part of
$\hat{\Delta}$ is
 described only by the wave number on the Fermi surface,
so we introduce $\hat{\boldsymbol{k}}= \boldsymbol{k}_F/|\boldsymbol{k}_F|$ with
$\boldsymbol{k}_F$ being the Fermi wave vector.
In this paper, we do not consider superconductors in which the
even- and odd-frequency pair potentials coexist.\cite{kusunose_prb}
When $\hat{\Delta}$ in Eq.\ (\ref{delta0}) is an even (odd) function of $\omega_n$,
the pair potential belongs to the even-frequency (odd-frequency) symmetry class $s_f=1$ ($s_f=-1$).
The conventional equal-time pair potential that does not depend on $\omega_n$, belongs to the even-frequency symmetry
class.
In what follows, in order to simplify the theory, we assume that the pairing interaction in Eqs.\ (\ref{delta_def1})
and (\ref{deltatilde_def1}) is spin-diagonal,
\begin{align}
V_{\alpha\beta;\gamma\delta}(\mathrm{x},\mathrm{x}')&=V_{\alpha,\beta}(\mathrm{x},\mathrm{x}')
\; \delta_{\alpha\gamma}\; \delta_{\beta\delta},\\
V_{\alpha,\beta}(\mathrm{x},\mathrm{x}')&=V_{\beta,\alpha}(\mathrm{x}',\mathrm{x}).
\end{align}

\subsection{Eilenberger equation}
Applying the standard gradient expansion,\cite{larkin2} we derive the
Eilenberger equation for quasiclassical Green functions.
We assume that the amplitude of the pair potential $\Delta$ is much smaller than
the Fermi energy $\mu_F$. The gradient expansion is justified for
$\xi_0 \gg \lambda_F$, where $\xi_0=v_F/(\pi \Delta)$ is the coherence length,
$v_F=k_F/m$ is the absolute value of the Fermi velocity,
and $\lambda_F$ is the Fermi wave length.
We first apply the Fourier transformation to the relative coordinate
of the Green functions,
\begin{align}
\hat{\mathfrak{G}}(\boldsymbol{r}_{12}, \boldsymbol{\rho}_{12}, \tau_{12})
=& \int \frac{d\boldsymbol{k}}{(2\pi )^d} T\sum_{\omega_n}
\hat{\mathfrak{G}}(\boldsymbol{r}_{12}, {\boldsymbol{k}}, i\omega_n)  \nonumber\\
&\times e^{i\boldsymbol{k}\cdot\boldsymbol{\rho}_{12}} e^{-i\omega_n \tau_{12}}.
\end{align}
The momentum integration is replaced by
\begin{align}
\int \frac{d\boldsymbol{k}}{(2\pi )^d} = N_0 \int \frac{d\hat{\boldsymbol{k}}}{S_d}
\int d\xi_k,\quad \xi_k=\frac{k^2}{2m}-\mu_F,
\end{align}
where $S_d$ is the full solid angle in $d$ dimensions and $N_0$ is the density of states
per spin at the Fermi level.
The quasiclassical Green functions are defined by
\begin{align}
\hat{g}(\boldsymbol{r}, \hat{\boldsymbol{k}}, i\omega_n)
=\frac{i}{\pi}\oint d\xi_k \hat{\mathfrak{G}}(\boldsymbol{r}, {\boldsymbol{k}}, i\omega_n),
\end{align}
where $\oint$ takes into account the contribution near the Fermi level.\cite{kopnin}
Other Green functions, $\underline{\hat{\mathfrak{G}}}$,
$\hat{\mathfrak{F}}$, and $\underline{\hat{\mathfrak{F}}}$, are transformed in the
same manner to the quasiclassical Green functions $\underline{\hat{g}}$,
$\hat{f}$, and $\underline{\hat{f}}$, respectively.
Such quasiclassical Green functions obey the $4\times 4$ Eilenberger equation
given by
\begin{align}
&i v_F \hat{\boldsymbol{k}} \cdot
\boldsymbol{\nabla}_{\boldsymbol{r}} \, \check{{g}}
+\left[ \check{H} +\check{\Sigma}, \check{g} \right]=0,\label{eilenberger_44_m}\\
&\check{H}
=\left[\begin{array}{cc}
\xi(\boldsymbol{r}, \hat{\boldsymbol{k}}, i\omega_n) \hat{\sigma}_0
&
\hat{\Delta}(\boldsymbol{r}, \hat{\boldsymbol{k}}, i\omega_n) \\
{\hat{\Delta}}^\ast(\boldsymbol{r}, -\hat{\boldsymbol{k}}, -i\omega_n) &
{\xi}^\ast(\boldsymbol{r}, -\hat{\boldsymbol{k}}, i\omega_n) \hat{\sigma}_0
\end{array}\right],
\label{hr_44_m} \\
&\check{g}
=\left[\begin{array}{cc}
\hat{g}(\boldsymbol{r}, \hat{\boldsymbol{k}}, i\omega_n) &
\hat{f}(\boldsymbol{r}, \hat{\boldsymbol{k}}, i\omega_n)\\
-\underline{\hat{f}}(\boldsymbol{r}, \hat{\boldsymbol{k}}, i\omega_n) &
-\underline{\hat{g}}(\boldsymbol{r}, \hat{\boldsymbol{k}}, i\omega_n)
\end{array}\right]\label{g_44_m},\\
&\xi=i\omega_n + ev_F \hat{\boldsymbol{k}}\cdot \boldsymbol{A}(\boldsymbol{r}), \label{xi_m}\\
&\check{\Sigma}=\frac{i}{2\tau_\mathrm{imp}} \left\langle \check{g}(\boldsymbol{r},\hat{\boldsymbol{k}},i\omega_n) \right\rangle_{\hat{\boldsymbol{k}}},\\
&\left\langle \check{g}(\boldsymbol{r},\hat{\boldsymbol{k}},i\omega_n) \right\rangle_{\hat{\boldsymbol{k}}}\equiv \int\frac{d\hat{\boldsymbol{k}}}{S_d} \check{g}(\boldsymbol{r},\hat{\boldsymbol{k}},i\omega_n),
\end{align}
where $\tau_\mathrm{imp}$
is the elastic mean free time.
The (2,1) component in Eq.\ (\ref{hr_44_m}) is represented by
\begin{align}
\hat{\Delta}^\ast(\boldsymbol{r},-\hat{\boldsymbol{k}}, -i\omega_n)
= s_f \hat{\Delta}^\ast(\boldsymbol{r},-\hat{\boldsymbol{k}}, i\omega_n).
\end{align}
The pair potential is related to the anomalous Green function,
\begin{align}
\hat{\Delta}(\boldsymbol{r},\hat{\boldsymbol{k}}, i\omega_n)
=&\int\frac{d\hat{\boldsymbol{k}'}}{S_d} T\sum_{\omega_m}
V_p(\hat{\boldsymbol{k}},\hat{\boldsymbol{k}}')
V_f(i\omega_n,i\omega_m)\notag\\
&\times \hat{f}(\boldsymbol{r},\hat{\boldsymbol{k}}', i\omega_m),\label{sceq_m}
\end{align}
where $V_p$ and $V_f$ are the potential representing the attractive interaction
between electrons.

To shorten further notations, we define
\begin{align}
\undertilde{X}(\boldsymbol{r}, \hat{\boldsymbol{k}}, i\omega_n) \equiv
{X}^\ast(\boldsymbol{r}, -\hat{\boldsymbol{k}}, i\omega_n)
\end{align}
for all functions of the Matsubara frequency.

From the symmetry of $\check{H}$,
\begin{align}
\check{T}_1\,  \undertilde{\check{H}} \,
\check{T}_1 &= \check{H} \quad (s_f=1),\\
i\check{T}_2 \, \undertilde{\check{H}} \,
\left(-i\check{T}_2 \right) &= \check{H}, \quad (s_f=-1),
\end{align}
we find
\begin{align}
\underline{\hat{f}}(\boldsymbol{r}, \hat{\boldsymbol{k}}, i\omega_n) &= s_f
\undertilde{\hat{f}}(\boldsymbol{r}, \hat{\boldsymbol{k}}, i\omega_n),\\
\underline{\hat{g}}(\boldsymbol{r}, \hat{\boldsymbol{k}}, i\omega_n) &=
\undertilde{\hat{g}}(\boldsymbol{r}, \hat{\boldsymbol{k}}, i\omega_n),
\end{align}
where we have introduced $4\times 4$ matrices,
\begin{equation}
\check{T}_1 = \left[ \begin{array}{cc} \hat{0} & \hat{\sigma}_0 \\ \hat{\sigma}_0 & \hat{0} \end{array} \right],\quad
\check{T}_2 = \left[ \begin{array}{cc} \hat{0} & -i\hat{\sigma}_0 \\ i\hat{\sigma}_0 & \hat{0} \end{array} \right].
\end{equation}
Thus, Eqs.\ (\ref{hr_44_m}) and (\ref{g_44_m}) can be written as
\begin{align}
\check{H}
&=\left[\begin{array}{cc}
\xi \hat{\sigma}_0
&
\hat{\Delta} \\
s_f \undertilde{\hat{\Delta}} &
\undertilde{\xi} \hat{\sigma}_0
\end{array}\right]_{(\boldsymbol{r}, \hat{\boldsymbol{k}}, i\omega_n)},
\label{hr_44_m2} \\
\check{g}
&=\left[\begin{array}{cc}
\hat{g} &
\hat{f}\\
-s_f\undertilde{\hat{f}} &
-\undertilde{\hat{g}}
\end{array}\right]_{(\boldsymbol{r}, \hat{\boldsymbol{k}}, i\omega_n)},
\label{g_44_m2}
\end{align}
with the normalization condition $\hat{g}^2- \hat{f}s_f \undertilde{\hat{f}}
=\hat{\sigma}_0$.
In the presence of spin-dependent potentials,
we need to add
\begin{align}
\hat{\xi}_{\textrm{spin}}=&-\boldsymbol{V}(\boldsymbol{r}) \cdot \hat{\boldsymbol{\sigma}}
- \boldsymbol{\lambda}(\boldsymbol{r}) \times \hat{\boldsymbol{\sigma}} \cdot \hat{\boldsymbol{k}}
\end{align}
to $\xi \hat{\sigma}_0$ and add
$\undertilde{\hat{\xi}}_{\textrm{spin}}$ to $\undertilde{\xi} \hat{\sigma}_0$ in Eq.\ (\ref{hr_44_m2}),
where $\boldsymbol{V}(\boldsymbol{r})$ and
$\boldsymbol{\lambda}(\boldsymbol{r})$ represent the exchange potential
and the spin-orbit coupling, respectively.
It is also possible to obtain the symmetry relationship between the Green functions,
\begin{align}
\hat{g}(\boldsymbol{r}, \hat{\boldsymbol{k}}, -i\omega_n)
&=- \hat{g}^\dagger(\boldsymbol{r}, \hat{\boldsymbol{k}}, i\omega_n),\\
\hat{f}^\mathrm{T}(\boldsymbol{r}, -\hat{\boldsymbol{k}}, -i\omega_n)
&= - {\hat{f}}(\boldsymbol{r}, \hat{\boldsymbol{k}}, i\omega_n), \label{fd_f}
\end{align}
by using the following symmetry of $\check{H}$:
\begin{align}
\check{T}_3 \check{H}^\dagger(\boldsymbol{r}, \hat{\boldsymbol{k}}, -i\omega_n)
 \check{T}_3 &= \check{H}(\boldsymbol{r}, \hat{\boldsymbol{k}}, i\omega_n), \quad (s_f=1),\\
\check{H}^\dagger(\boldsymbol{r}, \hat{\boldsymbol{k}}, -i\omega_n)
 &= \check{H}(\boldsymbol{r}, \hat{\boldsymbol{k}}, i\omega_n), \quad (s_f=-1).
\end{align}
Equation (\ref{fd_f}), where $\mathrm{T}$ denotes the matrix transposition, represents the Fermi-Dirac statistics of electrons.

When $\hat{\Delta}$ has only one spin component as
\begin{equation}
\hat{\Delta}(\boldsymbol{r},\hat{\boldsymbol{k}},i\omega_n)
={\Delta}(\boldsymbol{r},\hat{\boldsymbol{k}},i\omega_n)  i \hat{\sigma}_\nu \hat{\sigma}_2
e^{i\varphi},
\end{equation}
with $\nu$ being one of $0,1,2,3$,
it is possible to reduce the $4\times 4$ matrix equation to a $2 \times 2$ one.
Expressing the spin components of the Green functions as
\begin{align}
\hat{g}(\boldsymbol{r},\hat{\boldsymbol{k}},i\omega_n)
&=g(\boldsymbol{r},\hat{\boldsymbol{k}},i\omega_n)\hat{\sigma}_0,\\
\hat{f}(\boldsymbol{r},\hat{\boldsymbol{k}},i\omega_n)
&=f(\boldsymbol{r},\hat{\boldsymbol{k}},i\omega_n) \hat{\sigma}_\nu \hat{\sigma}_2,
\end{align}
we can write the Eilenberger equation in the clean limit as
\begin{align}
&i v_F \hat{\boldsymbol{k}} \cdot
\boldsymbol{\nabla}_{\boldsymbol{r}} \, \hat{{g}}
+\left[ \hat{H}, \hat{g} \right]=0,\label{eilenberger_22_m}\\
&\hat{H}
=\left[\begin{array}{cc}
\xi(\boldsymbol{r}, \hat{\boldsymbol{k}}, i\omega_n) &
i{\Delta}(\boldsymbol{r},\hat{\boldsymbol{k}},i\omega_n) \\
-is_s s_f \undertilde{\Delta}(\boldsymbol{r},\hat{\boldsymbol{k}},i\omega_n) &
\undertilde{\xi}(\boldsymbol{r}, \hat{\boldsymbol{k}}, i\omega_n)
\end{array}\right],\label{hr_m} \\
&\hat{g}
=\left[\begin{array}{cc}
{g}(\boldsymbol{r},\hat{\boldsymbol{k}},i\omega_n) &
{f}(\boldsymbol{r},\hat{\boldsymbol{k}},i\omega_n) \\
-s_s\; s_f\; \undertilde{f}(\boldsymbol{r},\hat{\boldsymbol{k}},i\omega_n) &
-g(\boldsymbol{r},\hat{\boldsymbol{k}},i\omega_n)
\end{array}\right].\label{g_22_m}
\end{align}
The pair potential obeys $s_s  \Delta(\boldsymbol{r},-\hat{\boldsymbol{k}},-i\omega_n) =-\Delta(\boldsymbol{r},\hat{\boldsymbol{k}},i\omega_n)$.
From the normalization condition, we also obtain
\begin{align}
&{g}(\boldsymbol{r},\hat{\boldsymbol{k}},i\omega_n)=
\undertilde{g}(\boldsymbol{r},\hat{\boldsymbol{k}},i\omega_n), \label{ggt}\\
&g^2(\boldsymbol{r},\hat{\boldsymbol{k}},i\omega_n) - s_s \; s_f \; f(\boldsymbol{r},\hat{\boldsymbol{k}},i\omega_n)
 \undertilde{f}(\boldsymbol{r},\hat{\boldsymbol{k}},i\omega_n)=1.
\end{align}
The components of the Green functions satisfy the following relations:
\begin{align}
g(\boldsymbol{r}, \hat{\boldsymbol{k}}, i\omega_n) &= -
g^\ast(\boldsymbol{r}, \hat{\boldsymbol{k}}, -i\omega_n), \label{gpgn}\\
s_s{f}(\boldsymbol{r},-\hat{\boldsymbol{k}},-i\omega_n)&=
-f(\boldsymbol{r},\hat{\boldsymbol{k}},i\omega_n).\label{far_m1}
\end{align}
In uniform superconductors, we obtain the solution as
\begin{equation}
\hat{g}(\hat{\boldsymbol{k}},i\omega_n)=\frac{1}{\Omega_n}
\left[ \begin{array}{cc} \omega_n & \Delta(\hat{\boldsymbol{k}},i\omega_n) \\
\Delta^\ast(\hat{\boldsymbol{k}},i\omega_n) & -\omega_n
\end{array}\right],
\end{equation}
with $\Omega_n=\sqrt{ \omega_n^2 + |\Delta(\hat{\boldsymbol{k}},i\omega_n) |^2}$
for both the even- and odd-frequency pair potentials.

\subsection{Real-energy representation}
Now we briefly discuss the Eilenberger equation in the real-energy representation.
For all functions in this representation, we define
\begin{equation}
\undertilde{X}(\boldsymbol{r}, \hat{\boldsymbol{k}}, \epsilon \pm i \delta) \equiv
{X}^\ast(\boldsymbol{r}, -\hat{\boldsymbol{k}}, -\epsilon \pm i \delta).
\end{equation}
The frequency symmetry of pair potential is described by
\begin{equation}
\hat{\Delta}(\boldsymbol{r},\hat{\boldsymbol{k}}, -\epsilon)
= s_f \hat{\Delta}(\boldsymbol{r},\hat{\boldsymbol{k}}, \epsilon).
\label{fsym_r}
\end{equation}
We note that the pair potential has no causality because it is the
potential in the mean-field theory of superconductivity.
Applying the analytic continuation in Eq.\ (\ref{sceq_m}),
we obtain
\begin{align}
\hat{\Delta}&(\boldsymbol{r},\hat{\boldsymbol{k}}, \epsilon)
=\int\frac{d\hat{\boldsymbol{k}'}}{S_d} \int \frac{d\epsilon'}{2\pi}
\tanh\left(\frac{\epsilon'}{2T}\right)
 V_p(\hat{\boldsymbol{k}},\hat{\boldsymbol{k}}')\notag\\
&\times V_f(\epsilon,\epsilon')
\left[\hat{f}^R(\boldsymbol{r},\hat{\boldsymbol{k}}', \epsilon')-
\hat{f}^A(\boldsymbol{r},\hat{\boldsymbol{k}}', \epsilon')\right].
\end{align}

The Eilenberger equation reads
\begin{align}
&i v_F \hat{\boldsymbol{k}} \cdot
\boldsymbol{\nabla}_{\boldsymbol{r}} \, \check{{g}}^{R(A)}
+\left[ \check{H}^{R(A)}, \check{g}^{R(A)} \right]=0,\label{eilenberger_44_r}\\
&\check{H}^{R(A)}
=\left[\begin{array}{cc}
\xi^{R(A)} \hat{\sigma}_0
&
\hat{\Delta} \\
\undertilde{\hat{\Delta}} &
\undertilde{\xi}^{R(A)} \hat{\sigma}_0
\end{array}\right]_{(\boldsymbol{r}, \hat{\boldsymbol{k}}, \epsilon)},
\label{hr_44_r} \\
&\check{g}^{R(A)}
=\left[\begin{array}{cc}
\hat{g}^{R(A)} &
\hat{f}^{R(A)}\\
-\undertilde{\hat{f}}^{R(A)} &
-\undertilde{\hat{g}}^{R(A)}
\end{array}\right]_{(\boldsymbol{r}, \hat{\boldsymbol{k}}, \epsilon)},
\label{g_44_r}\\
& \xi^{R(A)} = \epsilon \pm i\delta + ev_F \hat{\boldsymbol{k}}\cdot \boldsymbol{A}.
\end{align}
The advanced functions are related to the retarded ones as
\begin{equation}
\check{g}^A(\boldsymbol{r}, \hat{\boldsymbol{k}}, \epsilon) =
- \check{T}_3
\left[\check{g}^R(\boldsymbol{r}, \hat{\boldsymbol{k}}, \epsilon)\right]^\dagger
\check{T}_3.\label{grga_44}
\end{equation}
We also obtain the relation
\begin{equation}
\left[\hat{f}^A(\boldsymbol{r}, -\hat{\boldsymbol{k}}, -\epsilon)\right]^\mathrm{T} =
- \hat{f}^R(\boldsymbol{r}, \hat{\boldsymbol{k}}, \epsilon),
\end{equation}
which represents the Fermi-Dirac statistics of electrons.
In the real-energy representation, the Eilenberger equation
Eqs.\ (\ref{eilenberger_44_r})-(\ref{g_44_r})
and the symmetry relationship in Eq.\ (\ref{grga_44})
have the same form in the two cases: even-frequency pair potential
and the odd-frequency one.

\section{Inhomogeneous superconductors}

To analyze the symmetry of the subdominant component in inhomogeneous superconductors,
we begin our discussion with the $2 \times 2$ Eilenberger equation (\ref{eilenberger_22_m}).
In the absence of magnetic field (i.e., $\boldsymbol{A}=0$), it is possible to choose the gauge so that
the pair potential is real. As a result, all the Green functions in Eq.\ (\ref{g_22_m})
are real.
The Eilenberger equation can be decomposed into three equations,\cite{review}
\begin{align}
v_F \hat{\boldsymbol{k}}\cdot\nabla  g &= 2\Delta f_\mathrm{S}, \label{e_g}\\
 v_F \hat{\boldsymbol{k}}\cdot\nabla f_\mathrm{B} &=  - 2\omega_n f_\mathrm{S},
\label{e_fp}\\
 v_F \hat{\boldsymbol{k}}\cdot\nabla f_\mathrm{S} &= 2 (\Delta g -  \omega_n f_\mathrm{B} ),
\label{e_fm}\\
f_\mathrm{B}
=
\frac{1}{2}\left(f+s_p \undertilde{f} \right),\quad &
f_\mathrm{S}
=
\frac{1}{2}\left( f-s_p \undertilde{f}\right), \label{fbfs}
\end{align}
with the normalization condition $g^2+f s_p \undertilde{f}=
g^2 + f_{\mathrm{B}}^2 -f_{\mathrm{S}}^2=1$.
Here we omit the arguments of all the functions above,
(i.e., $\boldsymbol{r}, \hat{\boldsymbol{k}}, i\omega_n)$.
The functions $f_\mathrm{B}$ and $f_\mathrm{S}$ introduced in Eq.\ (\ref{fbfs}) can be interpreted as the bulk (dominant) and surface (subdominant) components of superconducting correlation, respectively.

In homogeneous superconductors, we obtain the solution as
\begin{equation}
g= {\omega_n}/{\Omega_n}, \quad
 f=s_p\undertilde{f}=
f_\mathrm{B}= {\Delta( \hat{\boldsymbol{k}}, i\omega_n)}
/{\Omega_n}
\end{equation}
and $f_\mathrm{S}=0$.
The weak coupling theory requires $\lim_{\omega_n\to\infty} \Delta(\hat{\boldsymbol{k}},i\omega_n)/\omega_n=0$ (which mean that at large frequencies the behavior is normal-metallic).
Parity, spin configuration and frequency symmetry of
$f_\mathrm{B}$ and those of the pair potential are identical
because they are linked to each other through the self-consistency equation.

Applying $v_F\hat{\boldsymbol{k}}\cdot \nabla$ to Eq.\ (\ref{e_fm}), we obtain
\begin{equation}
v_F^2 (\hat{\boldsymbol{k}}\cdot \nabla)^2 f_\mathrm{S} =
2g v_F (\hat{\boldsymbol{k}}\cdot \nabla \Delta)
+4 (\Delta^2+\omega_n^2) f_\mathrm{S}.
\end{equation}
The spatial derivative of the pair potential generates the $f_\mathrm{S}$ component
in inhomogeneous superconductors.
In what follows, we analyze the pairing symmetry of $f_\mathrm{S}$.
First of all, the spin configuration of $f_\mathrm{B}$ and that of $f_\mathrm{S}$
are the same because we do not consider any spin-dependent potentials.
From Eqs.\ (\ref{ggt}) and (\ref{gpgn}), we see that $g$ is an odd function of $\omega_n$
and has even parity when $g$ is real.
The left-hand side of Eq.\ (\ref{e_g}) is an odd function of $\omega_n$
and has odd parity because $\hat{\boldsymbol{k}}$ is an odd-parity function.
Thus, the frequency symmetry and parity of $f_\mathrm{S}$ are opposite to
those of $\Delta$.
Applying the same logic to Eq.\ (\ref{e_fp}),
we conclude that the frequency symmetry and parity of $f_\mathrm{S}$ are opposite to
those of $f_\mathrm{B}$. Therefore, $\Delta$ and $f_\mathrm{B}$
belong to the same symmetry class.

\begin{table}[t]
\begin{center}
\begin{ruledtabular}
\begin{tabular}{rcclc}
Frequency  & Spin & Parity& Magnetic response & \\
\hline
 $^{(a)}$ Even & Singlet & Even & Diamagnetic & Bulk \\
 Odd & Singlet & Odd & Paramagnetic & Induced\\
\hline
 $^{(b)}$ Even & Triplet & Odd & Diamagnetic & Bulk  \\
 Odd & Triplet & Even & Paramagnetic & Induced\\
\hline
 $^{(c)}$ Odd & Singlet & Odd & Diamagnetic & Bulk  \\
 Even & Singlet & Even & Paramagnetic& Induced\\
\hline
 $^{(d)}$ Odd & Triplet & Even & Diamagnetic & Bulk  \\
 Even & Triplet & Odd & Paramagnetic& Induced
\end{tabular}
\end{ruledtabular}
\end{center}
\caption[b]{
The classification of Cooper pairs in inhomogeneous superconductors.
In the absence of spin-dependent potentials, the spin state in the bulk and near an inhomogeneity is the same. At the same time, due to broken translational invariance, the spatial parity can change. This leads to changing of the frequency symmetry, in order to conform with the Pauli principle.
ESED states realized in metallic superconductors
and high-$T_c$ cuprates in~(a) have OSOP states as the subdominant component.
ETOD states realized in Sr$_2$RuO$_4$ and UPt$_3$ in~(b) have OTEP states as the subdominant component.
OSOD states in~(c) and OTED states in~(d) have never been confirmed in real materials.
The subdominant component of OSOD states in~(c) is ESEP states.
ETOP states appears as a subdominant component of OTED states in~(d).
}
\label{table1}
\end{table}

The results of the symmetry classification are summarized in Table~\ref{table1}.
The symmetry analysis of the anomalous Green functions in the presence of the vector potential $\boldsymbol{A}$
is presented in Appendix~\ref{sec:app:fBfS}.

Finally, we discuss magnetic properties of the two components $f_\mathrm{B}$ and $f_\mathrm{S}$.
The linear response of the electric current to the vector potential\cite{barash} is given by
\begin{align}
j_\mu(\boldsymbol{r})& = - \frac{e^2}{m} \mathcal{R}_{\mu\nu} A_\nu(\boldsymbol{r}),\\
\frac{\mathcal{R}_{\mu\nu}(\boldsymbol{r})}{n_e} &=
d \pi T\sum_{\omega_n}
 \left\langle \hat{k}_\mu\hat{k}_\nu \partial_{\omega_n}
 {g}(\boldsymbol{r},\hat{\boldsymbol{k}},i\omega_n) \right\rangle_{\hat{\boldsymbol{k}}}
 \notag \\
=d \pi T \sum_{\omega_n>0}&
 \left\langle \hat{k}_\mu\hat{k}_\nu
 \left( f_\mathrm{B}^2 - f_\mathrm{S}^2 \right)
 \partial_{\omega_n} \log\left( \frac{1+g}{1-g} \right)
\right\rangle_{\hat{\boldsymbol{k}}}
 \label{ns_mt},
\end{align}
where $\mathcal{R}(\boldsymbol{r})$ is the linear response tensor and $n_e$ is the electron density (see Appendix~\ref{sec:app:pairdensity} for details).
To reach the last equation, we have used the normalization condition in the Matsubara
representation.\cite{higashitani}
In simplest geometries, when $\mathcal R$ becomes diagonal, its elements are the so called \emph{pair densities}.
The sign of the pair density determines the type of magnetic response.
In uniform diamagnetic superconductors (i.e., when $f_\mathrm{S}=0$), the pair density
must be positive.
Therefore, the contribution of the $f_{\mathrm{B}}$ component in Eq.\ (\ref{ns_mt})
to the pair density is positive.
In inhomogeneous superconductors,
the induced component $f_{\mathrm{S}}$ gives negative contribution to the pair density as shown in Eq.\ (\ref{ns_mt}).
As a result, the subdominant Cooper pairs are paramagnetic (their contribution to the supercurrent is opposite to the conventional diamagnetic one).
The main conclusions of this paper are summarized in Table~\ref{table1} that classifies Cooper pairs into eight classes.

\section{Conventional Cooper pairs with paramagnetic response}
If we assume existence of odd-frequency spin-singlet $p_x$-wave (OSOD) superconductors,
then, according to the above analysis, the symmetry of $f_{\mathrm{S}}$ must be even-frequency spin-singlet even-parity.
At the same time, Table~\ref{table1} suggests that such conventional pairs have paramagnetic response (ESEP).

To confirm this general conclusion, we explicitly solve Eqs.\ (\ref{e_g})-(\ref{e_fm}) for
two particular cases.
Namely, we consider two types of $p_x$-wave superconductors
in two dimension and analyze the anomalous Green functions near the specularly reflecting surface at $x=0$.
The pair potentials given by
\begin{equation} \label{evenwoddw}
\Delta(x,\theta,i\omega_n) \times
\left\{
\begin{array}{ll} i\hat{\sigma}_1\hat{\sigma}_2 & \text{(ETOD),}\\
 i\hat{\sigma}_2 & \text{(OSOD),} \end{array}
\right.
\end{equation}
describe inhomogeneity introduced by the surface, while in the bulk we assume
\begin{equation}
\Delta(\infty,\theta,i\omega_n) \equiv \Delta(\theta,\omega_n) = \Delta_\infty Q(\omega_n) \cos\theta.
\end{equation}
Here the angle $\theta$ is counted from the normal to the surface (the $x$ axis).
We imply that the pair potential in Eq.\ (\ref{evenwoddw}) in the first case represents the equal-time spin-triplet
$p_x$-wave pair potential (i.e., ETOD state), while in the second case ---
the pair potential with odd-frequency spin-singlet $p_x$-wave symmetry (i.e., OSOD state).
Correspondingly, $Q(\omega_n)$ is equal to 1 in the ETOD case and is an odd (real) function of $\omega_n$ in the OSOD case.
It is known that the $p_x$-wave pair potential is suppressed at the surface.\cite{hara}
If we model this suppression by $\Delta(x,\theta,i\omega_n)=\Delta(\theta,\omega_n) \tanh(x/\xi)$ with $\xi=|v_F \cos\theta/\Delta(\theta,\omega_n)|$,
we can analytically find~\cite{schopohl} the Green functions satisfying the Eilenberger equations (\ref{e_g})-(\ref{e_fm}):
\begin{align}
g(x,\theta,\omega_n)&= \frac{\omega_n}{\Omega_n} +\frac{\Delta^2(\theta,\omega_n)}{2\omega_n\Omega_n}
\cosh^{-2}
\left(\frac{x}{\xi}\right),\label{gpx}\\
f_{\mathrm{B}}(x,\theta,\omega_n)&= \frac{\Delta(\theta,\omega_n)}{\Omega_n}\tanh\left(\frac{x}{\xi}\right),\\
f_{\mathrm{S}}(x,\theta,\omega_n)&= -\frac{\Delta^2(\theta,\omega_n)}{2\Omega_n}\cosh^{-2}
\left(\frac{x}{\xi}\right)\notag\\
& \times \left\{
\begin{array}{ll} \omega_n^{-1} & \text{for ETOD}\\
 |\omega_n|^{-1} & \text{for OSOD}, \end{array}\right.\label{fs}
\end{align}
with $\Omega_n=\sqrt{\omega_n^2+\Delta^2(\theta,\omega_n)}$. To obtain the last equation, we have used Eq.\ (\ref{far_m1}).

For OSOD pair potential,
the subdominant component $f_{\mathrm{S}}$ has the even-frequency spin-singlet symmetry.
In addition, Eq.\ (\ref{fs}) indicates that $f_{\mathrm{S}}$ is proportional to
$\cos^2(\theta)=( 1 + \cos 2\theta)/2 $ multiplied by
the even-parity functions.
Thus, $f_{\mathrm{S}}$ includes an $s$-wave component.
At the same time, according to Table~\ref{table1}, such conventional Cooper pairs are paramagnetic (ESEP).

\section{Discussion}

Experimentally, direct measurement of the paramagnetic subdominant components generated due to spatial inhomogeneity (see Table~\ref{table1}), is not an easy task. The main problem here is that in the general case the paramagnetic subdominant components are not spatially separated from the conventional diamagnetic ones (in terms of Table~\ref{table1}, one can say that near a spatial inhomogeneity the induced and bulk components coexist). At the same time, measurements allowing to probe quantities determined by the superfluid density (e.g., Meissner currents or magnetic susceptibility) are still well-suited for testing the magnetic properties of the induced components. The point is that the contribution of those components to the superfluid density is negative, so reduced superfluid density would be a sign of induced paramagnetic superconductivity.

The anomalous Meissner effect observed in films of high-$T_c$ superconductor with artificial internal surfaces (introduced by heavy-ion irradiation)\cite{walter} can be interpreted as the sign of the negative contribution to the pair density from subdominant pairing correlations. In the experiment, the penetration depth increased with decreasing temperature,\cite{walter} which suggests the decrease of the pair density at low temperature. In this case, according to our results, the OSOP state appears at the surface of the superconductor and decreases the pair density (in other words, appearance of the OSOP state is a manifestation of surface Andreev bound states formation\cite{tanuma}).
This picture is totally consistent with previous theoretical study on the paramagnetic instability
of small unconventional superconductors.\cite{suzuki}
Historically theoretical papers~\cite{yip,zare,fogelstrom,higashitani2,lofwander} have been tried to explain the anomalous Meissner effect
by the paramagnetic response of a quasiparticle at the surface Andreev bound states.
Even in conventional metallic superconductors,
vortex cores host the OSOP state in the clean limit.\cite{yokoyama} The presence of such paramagnetic Cooper pairs can be observed
through a large zero-energy peak in the quasiparticle density of states. These results also fully conform with our general classification.

Since surfaces in anisotropic superconductors can lead to appearance of subdominant components reducing the superfluid density, one can view this as suppression of superconductivity near a surface. This effect has a characteristic length scale given by the coherence length. Therefore, lateral confinement of a superconducting material by surfaces with distance between them of the order of the coherence length, can lead to complete suppression of superconductivity in the sample. This effect indeed exists, as shown theoretically in Refs.~\onlinecite{hara} and~\onlinecite{nagato}.

The variety of superconducting correlations discussed in the main part of this paper and classified in Table~\ref{table1} can be realized, at least in principle, in clean superconductors. The presence of impurities would lead to mixing different directions of quasiparticle motion, hence to isotropisation of the superconducting state and to suppression of anisotropic superconducting correlations. Diffusive motion near the surface (due to impurities or rough surface) is compatible only with the \textit{s}-wave components of the even-parity subdominant correlations [cases~(b) and~(c) in Table~\ref{table1}]. At the same time, all odd-parity subdominant components [cases~(a) and~(d)] are suppressed. Impurities of high concentration in the bulk (the dirty limit) would shrink the classification almost completely, allowing to consider only the \textit{s}-wave superconducting states in the bulk (even parity), which corresponds to cases~(a) and~(d).
At the same time, the corresponding subdominant correlations disappear because they belong to the odd-parity class.

Superconductors spontaneously break the continuous gauge symmetry
at the transition temperature. As a result, superconductors
acquire the diamagnetic response to magnetic field.
To form Cooper pairs, a superconductor
chooses three discrete symmetry options for Cooper pairing:
frequency symmetry, spin configuration, and parity.
This can be considered as
the secondary symmetry breaking, reflecting the material parameters such as
the lattice structure and the properties of bosons mediating the attractive
interactions of electrons.
When we describe the pair potential by $\Delta e^{i\varphi}$,
these symmetry options correspond to the inner
degrees of freedom of $\Delta$.
We have shown that the spatial inhomogeneity $\nabla \Delta$ generates the
subdominant pairing correlations whose frequency symmetry and parity are
opposite to those of the bulk state, as summarized in Table~\ref{table1}.
Therefore, the subdominant correlations behave as if recovering the broken
symmetries in the pairing options.

The equation $\nabla \varphi=0$ represents the phase coherence in the
uniform superconducting ground state.
The Nambu-Goldstone mode becomes massive and its typical excitation energy
is the plasma frequency because of the electron-electron interactions.
As a consequence, recovering the global gauge symmetry is suppressed in the
superconducting state.
Thus, the uniform phase-rigid ground state is stable, which is the basis
of success for the mean-field theory of superconductivity.
Table~\ref{table1} suggests that subdominant correlations are always
paramagnetic. This implies that the subdominant correlations attract magnetic field or
may generate self-induced magnetic field~\cite{mironov}.
The self-induced field encourages the gradient of phase
$\nabla \varphi$ which locally destroys the phase rigidity.
So, the subdominant correlations also play a role of recovering
the broken gauge symmetry.
In the ground state, nucleation of $\nabla \Delta$ requires
extrinsic triggers that bring inhomogeneity into the superconductor, such as
its own surfaces and interfaces to other materials, or vortices.
The subdominant correlations can be considered as a deformation of the uniform
phase-rigid ground state.
This would explain the ubiquitous presence of the paramagnetic Cooper pairs.

\section{Conclusions}
We have analyzed the symmetries and magnetic properties of Cooper pairs appearing
as a subdominant component of pairing correlations in inhomogeneous superconductors
by using the Eilenberger equation for the quasiclassical Green functions.
The spatial gradient of the pair potential generates the subdominant pairing correlations.
The frequency symmetry, parity and magnetic response of the subdominant component are
opposite to those of the dominant one in the bulk.
Therefore, the subdominant component can be interpreted as a deformation of the bulk superconducting state.
We conclude that the assumption of possibility of bulk odd-frequency superconducting states leads to a generalized classification of Cooper pairs into eight classes in terms of their pairing symmetries and magnetic properties.
Anomalous magnetic properties of subdominant components near surfaces can be probed by studying the Meissner effect (or, more generally, any effect sensitive to the superfluid density). Subdominant components generated in the vortex cores can be probed by the density of states measurement.

\begin{acknowledgments}
The authors are grateful to M.~V.\ Feigel'man, A.~A.\ Golubov, S.\ Higashitani, and H.\ Kusunose for helpful discussions.
This work was partially supported by the ``Topological Quantum Phenomena'' (Nos.~22103002, 22103005)
Grant-in Aid for
Scientific Research on Innovative Areas and KAKENHI (No.~26287069) from the Ministry of Education,
Culture, Sports, Science and Technology (MEXT) of Japan,
and by the Ministry of Education and Science of the Russian Federation (Grant No.~14Y.26.31.0007).
Ya.V.F.\ was supported in part by the program ``Quantum mesoscopic and disordered structures'' of the RAS,
and by the program ``5top100''.
\end{acknowledgments}

\appendix

\section{Pair density} \label{sec:app:pairdensity}

The pair density determines the response of superconductor to electromagnetic field.
Here we consider static magnetic field $\boldsymbol{H}=\nabla \times \boldsymbol{A}$.
In the Matsubara representation, the general expression for the electric current is given by
\begin{equation}
\boldsymbol{j}(\boldsymbol{r})= \frac{e v_F\pi N_0}{2i}
T\sum_{\omega_n}\int\frac{d\hat{\boldsymbol{k}}}{S_d}\hat{\boldsymbol{k}}
\Tr \left[\check{T}_3  \check{{g}}(\boldsymbol{r},\hat{\boldsymbol{k}},i\omega_n)\right],
\end{equation}
with $N_0$ being the density of states per spin at the Fermi level.
Here $\check{g}$ is the $4 \times 4$ Green function obeying Eq.\ (\ref{eilenberger_44_m})
with Eqs.\ (\ref{hr_44_m2}) and (\ref{g_44_m2}).
In the absence of spin-dependent potentials, we obtain
\begin{align}
\boldsymbol{j}(\boldsymbol{r})&= - {2ie v_F\pi N_0}
T\sum_{\omega_n}\int\frac{d\hat{\boldsymbol{k}}}{S_d}\hat{\boldsymbol{k}}
 g(\boldsymbol{r},\hat{\boldsymbol{k}},i\omega_n), \label{j_22_m}
\end{align}
for the single-spin-component pair potential, where $g$ is the scalar Green function
obeying Eqs.\ (\ref{eilenberger_22_m})-(\ref{g_22_m}).

In what follows, we find ${g}$ taking into account the linear response to $\boldsymbol{A}$.
In the Eilenberger equation in Eq.\ (\ref{eilenberger_22_m}) with Eq.\ (\ref{xi_m}),
the vector-potential term $ev_F \hat{\boldsymbol{k}}\cdot \boldsymbol{A}$ shifts the
Matsubara frequency.
Thus, the Green function can be expressed as
\begin{align}
g = g_0 + \partial_{\omega_n} g_0 (-iev_F) \hat{\boldsymbol{k}}\cdot \boldsymbol{A}
\end{align}
within the linear response, where $g_0$ is the Green function at $\boldsymbol{A}=0$.
Substituting this into Eq.\ (\ref{j_22_m}), we obtain (below we deal only with $g_0$ and omit the subscript for brevity)
\begin{align}
j_\mu(\boldsymbol{r}) &= - \frac{e^2}{m}\mathcal{R}_{\mu\nu}  A_\nu(\boldsymbol{r}),\\
\frac{\mathcal{R}_{\mu\nu}(\boldsymbol{r}) }{n_e} &=  d \pi T\sum_{\omega_n}
 \left\langle \hat{k}_\mu \hat{k}_\nu
 \partial_{\omega_n}
 {g}(\boldsymbol{r},\hat{\boldsymbol{k}}, i\omega_n) \right\rangle_{\hat{\boldsymbol{k}}}
 \label{ns_m}.
\end{align}
We have used the relation $ n_e= 2m v_F^2 N_0/d$, where $n_e$ is the electron density
and $d$ denotes the dimensionality of superconductor.
In the clean limit, anisotropic superconductivity can be realized, and then the linear response coefficient become a tensor due to dependence of $g$ on $\hat{\boldsymbol{k}}$.
In simplest geometries, when $\mathcal R$ becomes diagonal, its elements are the so called \emph{pair densities}.
To discuss magnetic properties of the anomalous Green function,
we rewrite the derivative of $g$ as\cite{higashitani}
\begin{equation}
 \partial_{\omega_n} g(\boldsymbol{r},\hat{\boldsymbol{k}},i\omega_n) = \frac{1}{2}
 \left( f_\mathrm{B}^2 - f_\mathrm{S}^2 \right)
 \partial_{\omega_n} \log\left( \frac{1+g}{1-g} \right),\label{ns_m2}
\end{equation}
where we have used the normalization condition in the Matsubara
representation, $g^2+ f_\mathrm{B}^2 - f_\mathrm{S}^2 =1$.
The $f_{\mathrm{B}}$ component and the $f_{\mathrm{S}}$ one contribute
to the response function in Eq.~(\ref{ns_m}) inversely to each other.

The electric current in the real-energy representation is also obtained
in the same way,
\begin{equation}
\boldsymbol{j}(\boldsymbol{r})=  -\frac{ev_F\pi N_0}{4}\!\!
\int\frac{d\epsilon}{2\pi}\int\frac{d\hat{\boldsymbol{k}}}{S_d}\hat{\boldsymbol{k}}
\Tr \left[\check{T}_3 \check{g}^K(\boldsymbol{r},\hat{\boldsymbol{k}},\epsilon)\right],
\end{equation}
where $\check{g}^K$ is the $4\times 4$ Keldysh Green function, which is given by
\begin{equation}
\check{g}^K(\boldsymbol{r},\hat{\boldsymbol{k}},\epsilon)
=\left[
\check{g}^R
-
\check{g}^A
\right]_{(\boldsymbol{r},\hat{\boldsymbol{k}},\epsilon)}
\tanh\left(\frac{\epsilon}{2T}\right)
\end{equation}
in equilibrium. In the absence of spin-dependent potentials, we obtain
\begin{equation}
\boldsymbol{j}(\boldsymbol{r})= - { 2ev_F\pi N_0}\!\!
\int\frac{d\epsilon}{2\pi}\!\!\int\frac{d\hat{\boldsymbol{k}}}{S_d}\hat{\boldsymbol{k}}
\Re \left[{g}^R(\boldsymbol{r},\hat{\boldsymbol{k}},\epsilon)\right],\label{j_eq}
\end{equation}
in the case of a single-spin-component pair potential.
Here $g^R$ is the normal Green function in the presence of the vector potential.
Within the linear response, we obtain
the response function as
\begin{equation}
\frac{\mathcal{R}_{\mu\nu}(\boldsymbol{r})}{n_e} = \frac{d}{2} \int_{-\infty}^{\infty}d\epsilon
\tanh\left(\frac{\epsilon}{2T}\right) \Re
 \left\langle \hat{k}_\mu\hat{k}_\nu \partial_\epsilon g^R(\boldsymbol{r},\hat{\boldsymbol{k}},\epsilon)
 \right\rangle_{\hat{\boldsymbol{k}}}.
\end{equation}
The real part of the normal Green function represents the local density of states and is an even function of
$\epsilon$.
Therefore $\partial_\epsilon \Re g^R$ is an odd function of $\epsilon$.

\section{Analysis of $f_\mathrm{B}$ and $f_\mathrm{S}$ in magnetic field}
 \label{sec:app:fBfS}

In the presence of magnetic field, Eqs.\ (\ref{e_g}), (\ref{e_fp}), and (\ref{e_fm})
should be generalized as
\begin{align}
v_F \hat{\boldsymbol{k}}\cdot\nabla  g &= 2 \left[ \Delta_c  f_{\mathrm{S}} - i\Delta_s
f_{\mathrm{B}} \right], \label{e_g2}\\
 v_F \hat{\boldsymbol{k}}\cdot\nabla f_{\mathrm{B}} &=  2\left[ -(\omega_n
-i ev_F \boldsymbol{k}\cdot \boldsymbol{A}) f_{\mathrm{S}}
+\Delta_s i g \right],
\label{e_fb2}\\
 v_F \hat{\boldsymbol{k}}\cdot\nabla f_{\mathrm{S}} &= 2 \left[
 \Delta_c g -  ( \omega_n -i ev_F \boldsymbol{k}\cdot \boldsymbol{A})  f_{\mathrm{B}}\right],
\label{e_fs2}\\
\Delta_c &= \Delta(\boldsymbol{r},\hat{\boldsymbol{k}},i\omega_n) \cos \varphi(\boldsymbol{r}),\\
 \Delta_s &= \Delta(\boldsymbol{r},\hat{\boldsymbol{k}},i\omega_n) \sin \varphi(\boldsymbol{r}),
\end{align}
where we have introduced coordinate-dependent phase of the pair potential,
$\varphi(\boldsymbol{r})$. The ground state is no longer uniform in this case.
The normal Green function can be represented as
\begin{equation}
g(\boldsymbol{r},\hat{\boldsymbol{k}},i\omega_n)=
g^r(\boldsymbol{r},\hat{\boldsymbol{k}},i\omega_n) +i
g^i(\boldsymbol{r},\hat{\boldsymbol{k}},i\omega_n),
\end{equation}
with $g^r$ and $g^i$ being real functions.
From Eqs.\ (\ref{ggt}) and (\ref{gpgn}),
we find that $g^r$ is an odd function of $\omega_n$ and has even parity.
On the other hand, $g^i$ is an even function of $\omega_n$ and has odd parity.
When we decompose the anomalous Green functions as
\begin{align}
f_{\mathrm{S}}(\boldsymbol{r},\hat{\boldsymbol{k}},i\omega_n) &=
f_{\mathrm{S}}^r(\boldsymbol{r},\hat{\boldsymbol{k}},i\omega_n) +i
f_{\mathrm{S}}^i(\boldsymbol{r},\hat{\boldsymbol{k}},i\omega_n),\\
f_{\mathrm{B}}(\boldsymbol{r},\hat{\boldsymbol{k}},i\omega_n) &=
f_{\mathrm{B}}^r(\boldsymbol{r},\hat{\boldsymbol{k}},i\omega_n) +i
f_{\mathrm{B}}^i(\boldsymbol{r},\hat{\boldsymbol{k}},i\omega_n),
\end{align}
we find that $\Delta$, $f_{\mathrm{B}}^r$, and $f_{\mathrm{S}}^i$ belong to the
same symmetry.
On the other hand,
the frequency symmetry and parity of $f_{\mathrm{B}}^i$ and $f_{\mathrm{S}}^r$
are opposite to those of $\Delta$.
For example, the real part of Eq.\ (\ref{e_fb2}) becomes
\begin{equation}
 v_F \hat{\boldsymbol{k}}\cdot\nabla f_{\mathrm{B}}^r =  2\left[
 -\omega_n f_{\mathrm{S}}^r
- ev_F \boldsymbol{k}\cdot \boldsymbol{A} f_{\mathrm{S}}^i
- \Delta_s  g^i \right].
\end{equation}
In the last term in the right-hand side, the frequency symmetry is the
same as that of $\Delta$, whereas the parity is opposite to $\Delta$.
Both the frequency symmetry and parity of $f_{\mathrm{B}}^r$ are
the same as those of $\Delta$ because of $\hat{\boldsymbol{k}}$ in the left-hand side.
In the same way, we find that $f_{\mathrm{S}}^i$ belongs to the same symmetry class as $\Delta$.
On the other hand, both the frequency symmetry and parity of $f_{\mathrm{S}}^r$
are opposite to those of $\Delta$.

\end{document}